\documentclass[showpacs,amsmath,amssymb,aps]{revtex4}                                                                                                                                                                                                                                                                                                                                                                                                                                                                                                                                                                                                                                                                                                                                                                                                                                                                                                                                                                                                                                                                                                                                                                                                                                                          
\usepackage{graphicx}
\usepackage{epstopdf}
\usepackage{amsmath}
\usepackage{bm}

\newcommand{\LL}{{\rm L}}
\newcommand{\RR}{{\rm R}}

\begin{document}

\title{Spontaneous Generation of Anisotropic Quasi-fermions}

\author{Kimihide Nishimura}
\email{kmdns@tune.ocn.ne.jp}
\affiliation{Nihon Uniform, Juso Motoimazato 1-4-21, Yodogawa-ku Osaka 532-0028 Japan}
%\address{Nihon Uniform, Juso Motoimazato 1-4-21, Yodogawa-ku Osaka 532-0028 Japan \email{kmdns@tune.ocn.ne.jp}}

\date{\today}

\begin{abstract}
We show that two types of relativistic quasi-fermions can emerge from spontaneous gauge and Lorentz violations of a chiral SU(2) model. In the terminology given in the previous  paper, the one consists of ``quasi-leptons", and the other consists of ``quasi-quarks" which are massless and have anisotropic dispersion relations characterized by constant vector potentials.
A low energy approximation shows that quasi fermions of the latter type are interpretable as collective excitations of the BCS-type vacuum, the Cooper pairs of which form vector mesons. The same approximation reproduces further a  generation structure similar to that of real quarks. 
\end{abstract}

%\subjectindex{12.10.-g, 12.38.-t, 74.20.-z, 11.30.Cp}

\maketitle

\section{Introduction}
This is a sequel of the papers \cite{KN1,KN2} proposing to describe leptons and quarks by quasi excitations emergent from spontaneous gauge and space-time symmetry violations.
The model has been constructed by the discipline that it should be based only on the fundamental principles still survivable even after excluding all the theoretical extensions or additional  hypotheses which have no inevitable necessities nor clear evidences.
In the context of local gauge field theories, this discipline leads us uniquely to a chiral SU(2) model in the following sense.

If the ultimate unified model of fermions exists, it has to generate all kinds of fermions from the least number of spinor fields. In four dimensional space-time, the group theory tells us that those are the left- and right-handed Weyl spinors. If they form one Dirac spinor, the model would not describe interactions other than the electromagnetism.   
On the other hand, if a left-handed Weyl spinor and a charge conjugate of the right-handed Weyl spinor form a left-handed doublet, then the SU(2) gauge interaction becomes allowable. 
This construction of a doublet would be favorable also in view of the explanation for the baryon asymmetry of the Universe, which requires interactions breaking the conservation of fermion number  \cite{Sakhalov,Trodden,KT}.
Due to the chiral anomaly, however, we could not further introduce abelian gauge interactions. As the result, we come to the  model with only a single chiral doublet and SU(2) gauge  interactions.

The first paper \cite{KN1} shows that if SU(2) gauge bosons become massive, a quasi fermion doublet interpretable as leptons actually emerges. Whereas this model seems to have also abilities to reproduce the structure of the standard theory as well as the baryon asymmetry of the Universe, the fact that quasi-leptons  emerge in a Lorentz-violating phase of vacuum ought to raise several questions. 

A serious one will be Lorentz invariance of the emergent theory.  
Though spontaneous Lorentz violation does not break the invariance under Lorentz transformations with respect to an observer, violations of relativity would possibly become manifest  in the dispersion relation of an emergent quasi fermion. 
It is known, however, that if a Lorentz-violating term in the Lagrangian has the same form as a constant electromagnetic potential, its magnitude is not restricted by experimental constraints  \cite{CK}.
Accordingly, if the dispersion relation of a quasi fermion have the ``quasi-relatitistic" form
\begin{equation}
\omega=\sqrt{(\bm{p}-\bm{\delta})^2+m^2}+\delta_0,
\label{QRDR}
\end{equation}
the Lorentz-violating constant 4-vector $\delta^\mu=(\delta_0,\bm{\delta})$ will cause no physical effect and therefore the emergent theory will be consistent with relativity, since $\delta^\mu$ is absorbable by a suitable local phase transformation for an effective Dirac field. 

This paper first proves that our model generates two types of quasi fermions with quasi-relativistic dispersion relations.
The one is essentially identical with the quasi-lepton doublet given in the first paper, and the other is composed of massless quasi fermions with anisotropic dispersion relations. We call here the second type of quasi fermions ``quasi-quarks", since the first paper suggests that the quasi fermions with anisotropic dispersion relations may be interpretable as quarks, though the emergence of them had not been confirmed. 

We next verify the emergence of quasi-quarks by simplifying our model in two ways.
One is performed by eliminating non-abelian gauge interactions, and using a perturbative method.
The other is by replacing further the SU(2) massive gauge interactions with four-fermion interactions, and using a variational method. 

The result from the first method shows that quasi-quarks are interpretable as collective excitations of the BCS-type vacuum \cite{BCS}, the Cooper pairs of which constitute vector mesons.
Then, we next try to reconstruct in turn quasi quarks from a trial vacuum composed of vectorial Cooper pairs with arbitrary coefficients. Two types of vacua are examined; one is constituted of Cooper pairs of vector meson type: $\vert{\rm  VM}\rangle$, and the other is of vector-dion type: $\vert{\rm  VD}\rangle$. We find that quasi-relativistic quasi-quarks are emergent only on the vacuum $\vert{\rm  VM}\rangle$, except for one singular situation.

We see that the four-fermion approximation can reproduce also the generation structure similar to that of quarks, since the extremum condition derived by the variational method generally allows several solutions. In our case, it gives maximally three quasi-quark generations. The dispersion relations are quasi relativistic for all  generations.  

Quasi fermions emergent from spontaneous gauge and Lorentz violations are not necessarily quasi-relativistic. We show this fact by constructing quasi fermions from the $\vert{\rm  VD}\rangle$ vacuum. Though they may not be interesting in view of Lorentz invariance, they exhibit a kind of mixing phenomenon, and may give some insights into phenomena of neutrino and quark mixings.

\section{Quasi-relativistic quasi-fermions}
The first paper \cite{KN1} assumes that quasi fermions emergent from gauge and Lorentz violations obey the following equation of motion
\begin{equation}
\begin{array}{lr}
(\bar{\sigma}^\mu i\partial_\mu-\bar{M})\Psi=0,& \bar{M}=\bar{\sigma}^\mu\displaystyle\frac{\rho_a}{2}m_a{}_\mu, 
\end{array}
\end{equation}
where the vacuum expectation values of SU(2) gauge potentials are expected to be the origin of $\bar{M}$.  
If they have quasi-relativistic dispersion relations (\ref{QRDR}), the identity:
\begin{equation}
|\bar{\sigma}\cdot p-\bar{M}|
=\left[(p-\delta_1)^2-m_1^2\right]\left[(p-\delta_2)^2-m_2^2\right],
\label{IDforDR}
\end{equation}
should hold  with respect to 4-momentum $p^\mu$, where $m_a^\mu=(0,\bm{m_a})$ for $a=1,2,3$ and $\delta_i^\mu=(\delta_i^0,\bm{\delta_i})$ are constant 4-vectors for $i=1,2$. From the coefficients of $p_0^3$ and $p_0^2$ of (\ref{IDforDR}) follow $\delta^\mu_1=-\delta^\mu_2=\delta^\mu$ and 
\begin{equation}
3\l_m^2/4=(\delta^0)^2+\bm{\delta}^2+(m_1^2+m_2^2)/2,
\label{Condition_A}
\end{equation}
where $\l_m$ is the mean root square of the edges of the parallelepiped formed with $\bm{m}_a$.
The coefficient of $p_0$ requires for the volume of the parallelepiped $V_m$,
\begin{equation}
V_m=2\delta_0[m_2^2-m_1^2+4\bm{p}\cdot\bm{\delta}],
\label{Condition_B}
\end{equation}
from which follows $\bm{\delta}=0$ or $\delta_0=0$, since  $V_m$ is independent of $\bm{p}$.
From the remaining terms follow three conditions:
\begin{equation}
\begin{array}{l}
\sum_a(\bm{e}_{\bm{p}}\cdot\bm{m}_a)^2
=4\left[(\delta^0)^2+(\bm{e}_{\bm{p}}\cdot\bm{\delta})^2\right],\\
(m_1^2-m_2^2)\bm{e}_{\bm{p}}\cdot\bm{\delta}=0,\\
(3\l_m^2/4)^2-3S_m^2/4=(m_1^2-\delta^2)(m_2^2-\delta^2),
\end{array}
\label{conditions_CDE}
\end{equation}
where $\bm{e}_{\bm{p}}=\bm{p}/|\bm{p}|$, and $S_m$ is the root mean square of the faces of the parallelepiped.

In the first case: $\bm{\delta}=0$, the first condition of (\ref{conditions_CDE}) requires $\bm{m_a}=m\bm{e}_a$ and $\delta^0=m/2$, where $m$ is a constant and $\bm{e}_a$ form some orthonormal basis. Then we have $\l_m^2=m^2$, $S_m^2=m^4$ and $V_m=m^3$. From the conditions (\ref{Condition_A}) and (\ref{Condition_B}) follow $m_1=0$ and $m_2=m$. Other conditions are automatically satisfied. Accordingly, we obtain
\begin{equation}
\begin{array}{cc}
\bar{M}=\displaystyle\frac{m}{2}\bm{\rho}\cdot O \cdot\bm{\sigma},&
p_0=\left\{
\begin{array}{l}
\pm|\bm{p}|+m/2,\\
\pm\sqrt{\bm{p}^2+m^2}-m/2,
\end{array}\right.
\end{array}
\label{QLD}
\end{equation}
where $O$ is an element of SO(3). The choice $O=1$ reduces (\ref{QLD}) to the form of a quasi-lepton doublet given in \cite{KN1}.

In the second case: $\delta^0=0$, on the other hand, from the second condition of (\ref{conditions_CDE}) follows $m_1^2=m_2^2=m^2$. Then from (\ref{Condition_A}) and the third condition of (\ref{conditions_CDE}) follows $S_m=0$, which implies that three vectors $\bm{m}_a$ have the same direction. Taking the first condition of (\ref{conditions_CDE}) into account, we obtain $\bm{m}_a=2\epsilon_a\bm{\delta}$ with some unit iso-vector $\bm{\epsilon}$, which gives $3\l_m^2/4=\bm{\delta}^2$, and therefore the masses of quasi-fermions vanish. As the result, we obtain 
\begin{equation}
\begin{array}{cc}
\bar{M}=\bm{\rho}\cdot\bm{\epsilon}\bm{\sigma}\cdot\bm{\delta},&
p_0=\left\{
\begin{array}{l}
\pm|\bm{p}-\bm{\delta}|,\\
\pm|\bm{p}+\bm{\delta}|,
\end{array}\right.
\end{array}
\label{QQD}
\end{equation}
which belongs to the  ``quasi-quark" doublets considered in \cite{KN1}.

\section{Quasi quarks}
\subsection{Perturbative method}
We prove in this section that quasi-quarks with the properties (\ref{QQD}) are actually emergent from either of the two simplified versions of our model. The Lagrangian of the one version consists of a chiral doublet with global SU(2) symmery:
\begin{equation}
{\cal L}={\cal L}_\Psi+{\cal L}_A
-\bm{j}^\mu\cdot\bm{A}_\mu ,
\label{Lag}
\end{equation}
\begin{equation}
\begin{array}{cc}
{\cal L}_\Psi=\Psi^\dagger\bar{\sigma}^\mu i\partial_\mu\Psi,
&
{\cal L}_A=-\frac{1}{4}\bm{F}^{\mu\nu}\cdot\bm{F}_{\mu\nu}
+\frac{1}{2}m^2_A\bm{A}^{\mu}\cdot\bm{A}_{\mu},
\end{array}
\end{equation}
\begin{equation}
\begin{array}{ccc}\bm{j}^\mu=g\Psi^\dagger\bar{\sigma}^\mu\frac{\bm{\rho}}{2}\Psi,&\Psi=\left(
\begin{array}{c}\varphi_1\\ \varphi_2\end{array}\right),&\bar{\sigma}^\mu=(1,-\bm{\sigma}),
\end{array}
\label{SU(2)current}
\end{equation}
where $\varphi_1$ and $\varphi_2 $ are left-handed Weyl spinors and $\bm{F}^{\mu\nu}$ contains no non-abelian part. 
We first divide the Lagrangian (\ref{Lag}) into a free part ${\cal L}_0={\cal L}_\Psi+{\cal L}_A-\Psi^\dagger\bar{M}\Psi$ and an interaction part ${\cal L}'={\cal L}-{\cal L}_0$.
Then the SU(2) current (\ref{SU(2)current}) develops a vacuum expectation value 
\begin{equation}
\begin{array}{cc}
\langle j^\mu_a\rangle=-{\rm Tr}g\bar{\sigma}^\mu\displaystyle\frac{\rho_a}{2} S(0),&
\left(S(0)=\displaystyle\int\frac{d^4p}{(2\pi)^4}
\frac{i}{\bar{\sigma}\cdot p-\bar{M}}\right),
\end{array}
\label{S(0)}
\end{equation}
which in turn generates an extra self-enegy $\Sigma$ for a chiral doublet:
\begin{equation}
\Sigma=g\bar{\sigma}^\mu\displaystyle\frac{\rho_a}{2}
\frac{\langle j^\mu_a\rangle}{m_A^2}.
\label{SCC}
\end{equation}
If $m_a^\mu=2\epsilon_a\delta^\mu$ satisfies the self consistency condition $\Sigma-\bar{M}=0$, then quasi-quarks can be emergent. 
 In this case, we find $S(0)=-\Gamma_QM$ with
\begin{equation}
\begin{array}{cccc}
M=\sigma^\mu\displaystyle\frac{\rho_a}{2}m_a{}_\mu
=\bm{\rho}\cdot\bm{\epsilon}\sigma\cdot\delta,&
\sigma^\mu=(1, \bm{\sigma}),
&\Gamma_Q=\displaystyle\frac{k_1}{2}-\frac{\bm{\delta}^2}{60\pi^2},&
k_1=\displaystyle\int\frac{d^4p}{(2\pi)^4}\frac{i}{p^2+i\epsilon},
\end{array}
\label{S-estimation}
\end{equation}
as shown in the Appendix.
The self consistency condition reduces to $g^2\Gamma_Q/m_A^2=1$, from which we obtain 
\begin{equation}
\begin{array}{cc}
\displaystyle\frac{|\bm{\delta}|}{m_A}=\frac{2\pi\sqrt{15}}{g}\sqrt{\xi^2-1},
&\left(\displaystyle\xi=\sqrt{\frac{g^2k_1}{2m_A^2}}
=\frac{g\Lambda}{4\pi m_A}\right),
\end{array}
\end{equation}
where $\Lambda$ is a 3-momemtum cut off. Then quasi-quarks can be emergent for $\xi>1$.

We can further reinterpret these quasi-quarks as collective excitations of the vacuum $\vert{\rm  VM}\rangle$.
For that purpose, we introduce annihilation operators $q_{1\bm{p}}$ and $q_{2\bm{p}}$ for quasi-quarks with energies 
$\omega_{\bm{p}-}=|\bm{p}-\bm{\delta}|$ and $
\omega_{\bm{p}+}=|\bm{p}-\bm{\delta}|$, while $\bar{q}_{1\bm{p}}$ and $\bar{q}_{2\bm{p}}$ for quasi-anti-quarks with energies 
$\omega_{\bm{p}+}$ and $
\omega_{\bm{p}-}$, respectively.
Expressing  the Schr\"odinger representation of $\Psi$ in terms of quasi-quark operators, we have 
\begin{equation}
\Psi(\bm{x})=\sum_{\bm{p}}
\left[\begin{array}{c}
q_{1\bm{p}}\chi_+\varphi_{\bm{p}-\bm{\delta}\LL}
+\bar{q}^\dagger_{1-\bm{p}}\chi_+
\varphi_{\bm{p}-\bm{\delta}\RR}\\
q_{2\bm{p}}\chi_-\varphi_{\bm{p}+\bm{\delta}\LL}
+\bar{q}^\dagger_{2-\bm{p}}\chi_-
\varphi_{\bm{p}+\bm{\delta}\RR}
\end{array}
\right]
\frac{e^{i\bm{p}\cdot\bm{x}}}{\sqrt{V}},
\label{QQPsi}
\end{equation}
where $\bm{\rho}\cdot\bm{\epsilon}\chi_\pm=\pm\chi_\pm$,  while $\varphi_{\bm{k}\RR}$ and $\varphi_{\bm{k}\LL}$ denote the left-handed and the right-handed helicity eigenstates with respect to momentum $\bm{k}$, respectively.

If the annihilation operators for primary fermions: $a_{\bm{p}\LL}$, $b_{\bm{p}\LL}$, $a_{\bm{p}\RR}$, $b_{\bm{p}\RR}$ are defined by
\begin{equation}
\begin{array}{ll}
\varphi_1(\bm{x})=\sum_{\bm{p}}(a_{\bm{p}\LL}\varphi_{\bm{p}\LL}+b_{-\bm{p}\RR}^\dagger \varphi_{\bm{p}\RR})\displaystyle\frac{e^{i\bm{p}\cdot\bm{x}}}{\sqrt{V}},&
\varphi_2(\bm{x})=\sum_{\bm{p}}(b_{\bm{p}\LL}\varphi_{\bm{p}\LL}+a_{-\bm{p}\RR}^\dagger \varphi_{\bm{p}\RR})\displaystyle\frac{e^{i\bm{p}\cdot\bm{x}}}{\sqrt{V}},
\end{array}
\label{aabb}
\end{equation}
the comparison between (\ref{QQPsi}) and (\ref{aabb}) gives 
\begin{equation}
\begin{array}{ll}
\left\{\begin{array}{l}
q_{1\bm{p}}=\alpha_{1\bm{p}}a'_{\bm{p}\LL}+\beta_{1\bm{p}}{b'}_{-\bm{p}\RR}^\dagger\\
q_{2\bm{p}}=\alpha_{2\bm{p}}b'_{\bm{p}\LL}+\beta_{2\bm{p}}{a'}_{-\bm{p}\RR}^\dagger
\end{array}\right.
,&
\left\{\begin{array}{l}
\bar{q}_{1\bm{p}}=\alpha_{1-\bm{p}}b'_{\bm{p}\RR}-\beta_{1-\bm{p}}{a'}_{-\bm{p}\LL}^\dagger\\
\bar{q}_{2\bm{p}}=\alpha_{2-\bm{p}}a'_{\bm{p}\RR}-\beta_{2-\bm{p}}{b'}_{-\bm{p}\LL}^\dagger
\end{array}\right.
\end{array},
\label{BT}
\end{equation}
where 
\begin{equation}
\begin{array}{ll}
a'_{\bm{p}\LL}=\chi_+^\dagger\left(\begin{array}{c}
a_{\bm{p}\LL}\\b_{\bm{p}\LL}
\end{array}\right),&
b'_{\bm{p}\LL}=\chi_-^\dagger\left(\begin{array}{c}
a_{\bm{p}\LL}\\b_{\bm{p}\LL}
\end{array}\right),\\
{b'}^\dagger_{-\bm{p}\RR}=\chi_+^\dagger
\left(\begin{array}{c}
b^\dagger_{-\bm{p}\RR}\\a^\dagger_{-\bm{p}\RR}
\end{array}\right),&
{a'}^\dagger_{-\bm{p}\RR}=\chi_-^\dagger
\left(\begin{array}{c}
b^\dagger_{-\bm{p}\RR}\\a^\dagger_{-\bm{p}\RR}
\end{array}\right),
\end{array}
\end{equation}
\begin{equation}
\begin{array}{cc}
\left\{\begin{array}{c}
\alpha_{1\bm{p}}=\varphi_{\bm{p}-\bm{\delta}\LL}^\dagger\varphi_{\bm{p}\LL},\\
\beta_{1\bm{p}}=\varphi_{\bm{p}-\bm{\delta}\LL}^\dagger\varphi_{\bm{p}\RR},
\end{array}\right.
&
\left\{\begin{array}{c}
\alpha_{2\bm{p}}=\varphi_{\bm{p}+\bm{\delta}\RR}^\dagger\varphi_{\bm{p}\LL},\\
\beta_{2\bm{p}}=\varphi_{\bm{p}+\bm{\delta}\RR}^\dagger\varphi_{\bm{p}\RR}.
\end{array}\right.
\end{array}
\end{equation}
The complex coefficients $\alpha_{i\bm{p}}$ and $\beta_{i\bm{p}}$ satisfy the conditions 
\begin{equation}
\begin{array}{cr}
|\alpha_{i\bm{p}}|^2+|\beta_{i\bm{p}}|^2=1,&(i=1,2).
\end{array}
\label{Constraints}
\end{equation}
Then we see that quasi-quark operators $(q_{i\bm{p}}, \bar{q}_{i\bm{p}})$ annihilate the vacuum: 
\begin{equation}
|{\rm VM}\rangle=\prod_{\bm{p}}\left(\alpha_{1\bm{p}}+\beta_{1\bm{p}}(a'_{\bm{p}\LL}b'_{-\bm{p}\RR})^\dagger\right)\left(\alpha_{2\bm{p}}+\beta_{2\bm{p}}(b'_{\bm{p}\LL}a'_{-\bm{p}\RR})^\dagger\right)|0\rangle,
\label{VMV}
\end{equation}
which implies that the quasi-quarks are collective excitations of the vacuum $|{\rm VM}\rangle$.
\subsection{Variational method}
We next consider as the second approximation the following Hamiltonian with the four-fermion interactions:
\begin{equation}
H=\int d^3x\left[\Psi^\dagger i\bm{\sigma}\cdot\bm{\nabla}\Psi
+KV\Psi^\dagger\bar{\sigma}^\mu\frac{\rho_a}{2}\Psi\Psi^\dagger\bar{\sigma}_\mu\frac{\rho_a}{2}\Psi\right], 
\label{Ham}
\end{equation}
where the constant $K$ with mass dimension one is defined by $K=g^2/2m_A^2V$. 
We seek for the ground state of (\ref{Ham}) by the variational method. A trial wave function is prepared by regarding the coefficients $\alpha_{i\bm{p}}$ and $\beta_{i\bm{p}}$ in (\ref{VMV}) as arbitrary parameters satisfying the conditions (\ref{Constraints}).
Expressed in terms of 2-spinors, (\ref{Ham}) is rewritten by $H=H+H'$, where
\begin{equation}
\begin{array}{ll}
H_0=&\int d^3x\left[\varphi_1^\dagger i\bm{\sigma}\cdot\bm{\nabla}\varphi_1+\varphi_2^\dagger i\bm{\sigma}\cdot\bm{\nabla}\varphi_2\right],
\\
H'=&\int d^3x\left[\displaystyle\frac{1}{2}\left(\varphi_1^\dagger\bar{\sigma}^\mu\varphi_2\varphi_2^\dagger\bar{\sigma}_\mu\varphi_1
+\varphi_2^\dagger\bar{\sigma}^\mu\varphi_1\varphi_1^\dagger\bar{\sigma}_\mu\varphi_2\right)
+\displaystyle\frac{1}{4}\left(\varphi_1^\dagger\bar{\sigma}^\mu\varphi_1-\varphi_2^\dagger\bar{\sigma}^\mu\varphi_2\right)\left(\varphi_1^\dagger\bar{\sigma}_\mu\varphi_1-\varphi_2^\dagger\bar{\sigma}_\mu\varphi_2\right)\right].
\end{array}
\label{H_0&H'}
\end{equation}
The left-handed Weyl spinors $\varphi_1$ and $\varphi_2$ are also understood as time-independent Schr\"{o}dinger operators. Instead of (\ref{aabb}), we take another phase convention:
\begin{equation}
\begin{array}{ll}
\varphi_1(\bm{x})=\sum_{\bm{p}}(a_{\bm{p}\LL}e^{i\bm{p}\cdot\bm{x}}+b_{\bm{p}\RR}^\dagger e^{-i\bm{p}\cdot\bm{x}})\displaystyle\frac{\varphi_{\bm{p}\LL}}{\sqrt{V}},&
\varphi_2(\bm{x})=\sum_{\bm{p}}(b_{\bm{p}\LL}e^{i\bm{p}\cdot\bm{x}}+a_{\bm{p}\RR}^\dagger e^{-i\bm{p}\cdot\bm{x}})\displaystyle\frac{\varphi_{\bm{p}\LL}}{\sqrt{V}},
\end{array}
\label{aabb2}
\end{equation}
and assume further $a'_{\bm{p}\LL}=a_{\bm{p}\LL}$, $b'_{\bm{p}\LL}=b_{\bm{p}\LL}$, $a'_{\bm{p}\RR}=a_{\bm{p}\RR}$, $b'_{\bm{p}\RR}=b_{\bm{p}\RR}$, $\bm{\epsilon}=(0,0,1)$, for simplicity. Then the energy of the vacuum  
$W=\langle{\rm VM}| H|{\rm VM}\rangle$ is estimated as 
\begin{equation}
\begin{array}{ll}
W&=\displaystyle\sum_{\bm{p}}2|\bm{p}|(|\beta_{1\bm{p}}|^2+|\beta_{2\bm{p}}|^2)\\
&+K\displaystyle\sum_{\bm{p},\bm{q}}\left[
\begin{array}{l}
-3/2\\
-2\bm{e}_{\bm{p}}\cdot\bm{e}_{\bm{q}}(|\beta_{1\bm{p}}|^2|\beta_{1\bm{q}}|^2+|\beta_{2\bm{p}}|^2|\beta_{2\bm{q}}|^2+|\beta_{1\bm{p}}|^2|\beta_{2\bm{q}}|^2)\\
-i\bm{e}_{\bm{p}}\cdot\bm{e}^+_{\bm{q}}|\beta_{1\bm{p}}|^2(2\alpha_{1\bm{q}}\beta^*_{1\bm{q}}+2\alpha^*_{1-\bm{q}}\beta_{1-\bm{q}}+\alpha_{2\bm{q}}\beta^*_{2\bm{q}}+\alpha^*_{2-\bm{q}}\beta_{2-\bm{q}})\\
-i\bm{e}_{\bm{p}}\cdot\bm{e}^+_{\bm{q}}|\beta_{2\bm{p}}|^2(\alpha_{1\bm{q}}\beta^*_{1\bm{q}}+\alpha^*_{1-\bm{q}}\beta_{1-\bm{q}}+2\alpha_{2\bm{q}}\beta^*_{2\bm{q}}+2\alpha^*_{2-\bm{q}}\beta_{2-\bm{q}})\\
+\frac{1}{2}\bm{e}^+_{\bm{p}}\cdot\bm{e}^+_{\bm{q}}
\left\{
\begin{array}{r}
(\alpha_{1\bm{p}}\beta^*_{1\bm{p}}+\alpha^*_{1-\bm{p}}\beta_{1-\bm{p}})
(\alpha_{1\bm{q}}\beta^*_{1\bm{q}}+\alpha^*_{1-\bm{q}}\beta_{1-\bm{q}})\\
+(\alpha_{2\bm{p}}\beta^*_{2\bm{p}}+\alpha^*_{2-\bm{p}}\beta_{2-\bm{p}})
(\alpha_{2\bm{q}}\beta^*_{2\bm{q}}+\alpha^*_{2-\bm{q}}\beta_{2-\bm{q}})\\
+(\alpha_{1\bm{p}}\beta^*_{1\bm{p}}+\alpha^*_{1-\bm{p}}\beta_{1-\bm{p}})
(\alpha_{2\bm{q}}\beta^*_{2\bm{q}}+\alpha^*_{2-\bm{q}}\beta_{2-\bm{q}})
\end{array}\right\}
\end{array}\right],
\end{array}
\label{VacuumEnergy}
\end{equation}
where $\bm{e}^\pm_{\bm{p}}=\bm{e}^1_{\bm{p}}\pm i\bm{e}^2_{\bm{p}}$ and $\bm{e}_{\bm{p}}=\bm{e}^3_{\bm{p}}$ in terms of the orthonormal basis vectors $\bm{e}^a_{\bm{p}}$. The explicit representations of $\bm{e}^a_{\bm{p}}$ are given by
\begin{equation}
\left\{\begin{array}{lcr} 
\bm{e}^1_{\bm{p}}=(-\sin\phi, &\cos\phi,&0),\\
\bm{e}^2_{\bm{p}}=(-\cos\theta\cos\phi, &-\cos\theta\sin\phi,&\sin\theta),\\
\bm{e}^3_{\bm{p}}=(\sin\theta\cos\phi, &\sin\theta\sin\phi,&\cos\theta),
\end{array}\right.
\label{orthnbasis}
\end{equation}
where $(\theta, \phi)$ are the polar coordinates of momentum $\bm{p}$. The terms $|\alpha_{i\bm{p}}|^2$ have been eliminated from (\ref{VacuumEnergy}) by using the constraints (\ref{Constraints}).

In order to find the ground state of the system,  we minimize $W$ with respect to the coefficients $\alpha_{i\bm{p}}$, $\beta_{i\bm{p}}$, $\alpha_{i\bm{p}}^*$ and $\beta_{i\bm{p}}^*$ under the constraints (\ref{Constraints}). Introducing Lagrange multipliers $\lambda_i$, we have a variational equation
\begin{equation}
\delta \left(W-\sum_{i=1}^2\sum_{\bm{p}}\lambda_{i\bm{p}}\left(|\alpha_{i\bm{p}}|^2+|\beta_{i\bm{p}}|^2-1\right)\right)=0,
\label{VE}
\end{equation}
which can be written in the form
\begin{equation}
\left\{\begin{array}{ll} 
\displaystyle\frac{\partial W}{\partial\alpha_{i\bm{p}}^*}&=A_{i\bm{p}}\alpha_{i\bm{p}}+B_{i\bm{p}}\beta_{i\bm{p}}=\lambda_{i\bm{p}}\alpha_{i\bm{p}},\\
\displaystyle\frac{\partial W}{\partial\beta_{i\bm{p}}^*}&=C_{i\bm{p}}\alpha_{i\bm{p}}+D_{i\bm{p}}\beta_{i\bm{p}}=\lambda_{i\bm{p}}\beta_{i\bm{p}}.\\
\end{array}\right.
\label{ABCD}
\end{equation}
Variations with respect to $\alpha_{i\bm{p}}$ and $\beta_{i\bm{p}}$ lead no new equations.
Since $A_{i\bm{p}}=0$, $C_{i\bm{p}}=B_{i\bm{p}}^*$, and that $D_{i\bm{p}}$ are real and contain an additive term $2|\bm{p}|$, we can introduce new parameters $m_{i\bm{p}}$ and $\delta_{i\bm{p}}$ by
\begin{equation}
\begin{array}{cc} 
m_{i\bm{p}}=-C_{i\bm{p}},&\delta_{i\bm{p}}=-|\bm{p}|+D_{i\bm{p}}/2.
\end{array}
\label{m and delta}
\end{equation}
The mass parameters $m_{i\bm{p}}$ are complex and the potential parameters $\delta_{i\bm{p}}$ are real. They depend on the direction of momentum $\bm{p}$, but not on the magnitude $|\bm{p}|$.

The Lagrange multipliers $\lambda_{i\bm{p}}$ should be  solutions of the eigen-equation: 
\begin{equation}
\left|\begin{array}{cc} 
-\lambda_{i\bm{p}}&-m_{i\bm{p}}^*\\
-m_{i\bm{p}}&2(|\bm{p}|+\delta_{i\bm{p}})-\lambda_{i\bm{p}}
\end{array}\right|=0,
\label{eigenequations}
\end{equation}
and we find two solutions
\begin{equation}
\begin{array}{cc} 
\lambda_{i\bm{p}}=|\bm{p}|+\delta_{i\bm{p}}\pm\omega_{i\bm{p}}, 
&\omega_{i\bm{p}}=\sqrt{(|\bm{p}|+\delta_{i\bm{p}})^2+|m_{i\bm{p}}|^2}.
\end{array}
\label{lambdas}
\end{equation}
The introduction of $m_{i\bm{p}}$ and $\delta_{i\bm{p}}$ rewrites the coefficients $\alpha_{i\bm{p}}$ and $\beta_{i\bm{p}}$ as 
\begin{equation}
\begin{array}{cc} 
\alpha_{i\bm{p}}=\sqrt{\displaystyle\frac{1}{2}\left(1\mp\frac{|\bm{p}|+\delta_{i\bm{p}}}{\omega_{i\bm{p}}}\right)},&\beta_{i\bm{p}}=\displaystyle\mp\frac{m_{i\bm{p}}}{|m_{i\bm{p}}|}\sqrt{\frac{1}{2}\left(1\pm\frac{|\bm{p}|+\delta_{i\bm{p}}}{\omega_{i\bm{p}}}\right)},
\end{array}
\label{alphabeta}
\end{equation}
in the phase convention: $\alpha^*_{i\bm{p}}=\alpha_{i\bm{p}}$.
The selection of the lower signs in (\ref{alphabeta}) minimizes 
 $W$, since otherwise $|\beta_{i\bm{p}}|\rightarrow1$ when $|\bm{p}|\rightarrow\infty$, which implies that high-momentum fermions would dominate the vacuum energy (\ref{VacuumEnergy}).

The equations (\ref{ABCD}) then give the self consistency equations for $m_{i\bm{p}}$ and $\delta_{i\bm{p}}$ :
\begin{equation}
\left\{\begin{array}{ll} 
m_{1\bm{p}}=\displaystyle K\sum_{\bm{p}}
&\quad i\bm{e}_{\bm{p}}^+\cdot\bm{e}_{\bm{q}}
\left(2|\beta_{1\bm{q}}|^2+|\beta_{2\bm{q}}|^2\right)\\
&-\frac{1}{2}\bm{e}^+_{\bm{p}}\cdot\bm{e}^+_{\bm{q}}
\left(2\alpha_{1\bm{q}}\beta^*_{1\bm{q}}+2\alpha^*_{1-\bm{q}}\beta_{1-\bm{q}}+\alpha_{2\bm{q}}\beta^*_{2\bm{q}}+\alpha^*_{2-\bm{q}}\beta_{2-\bm{q}}\right),
\\
\delta_{1\bm{p}}=\displaystyle K\sum_{\bm{p}}
&-\bm{e}_{\bm{p}}\cdot\bm{e}_{\bm{q}}\left(2|\beta_{1\bm{q}}|^2+|\beta_{2\bm{q}}|^2\right)\\
&-\frac{i}{2}\bm{e}_{\bm{p}}\cdot\bm{e}^+_{\bm{q}}
\left(2\alpha_{1\bm{q}}\beta^*_{1\bm{q}}+2\alpha^*_{1-\bm{q}}\beta_{1-\bm{q}}+\alpha_{2\bm{q}}\beta^*_{2\bm{q}}+\alpha^*_{2-\bm{q}}\beta_{2-\bm{q}}\right),
\end{array}\right.
\label{M1&D1}
\end{equation}
\begin{equation}
\left\{\begin{array}{ll} 
m_{2\bm{p}}=\displaystyle K\sum_{\bm{p}}
&\quad i\bm{e}_{\bm{p}}^+\cdot\bm{e}_{\bm{q}}
\left(|\beta_{1\bm{q}}|^2+2|\beta_{2\bm{q}}|^2\right)\\
&-\frac{1}{2}\bm{e}^+_{\bm{p}}\cdot\bm{e}^+_{\bm{q}}
\left(\alpha_{1\bm{q}}\beta^*_{1\bm{q}}+\alpha^*_{1-\bm{q}}\beta_{1-\bm{q}}+2\alpha_{2\bm{q}}\beta^*_{2\bm{q}}+2\alpha^*_{2-\bm{q}}\beta_{2-\bm{q}}\right),
\\
\delta_{2\bm{p}}=\displaystyle K\sum_{\bm{p}}
&-\bm{e}_{\bm{p}}\cdot\bm{e}_{\bm{q}}\left(|\beta_{1\bm{q}}|^2+2|\beta_{2\bm{q}}|^2\right)\\
&-\frac{i}{2}\bm{e}_{\bm{p}}\cdot\bm{e}^+_{\bm{q}}
\left(\alpha_{1\bm{q}}\beta^*_{1\bm{q}}+\alpha^*_{1-\bm{q}}\beta_{1-\bm{q}}+2\alpha_{2\bm{q}}\beta^*_{2\bm{q}}+2\alpha^*_{2-\bm{q}}\beta_{2-\bm{q}}\right).
\end{array}\right.
\label{M2&D2}
\end{equation}
We see from (\ref{M1&D1}) and (\ref{M2&D2}) that  $m_{i\bm{p}}$ and $\delta_{i\bm{p}}$ are expressible in the forms : $m_{i\bm{p}}= i\bm{e}_{\bm{p}}^+\cdot\bm{m}_i$ and $\delta_{i\bm{p}}= \bm{e}_{\bm{p}}\cdot\bm{\delta}_i$, where $\bm{m}_i$ and $\bm{\delta}_i$ are real constant 3-vectors.
Then we further see that $\bm{m}_i=-\bm{\delta}_i$, and consequently, $\omega_{i\bm{p}}=|\bm{p}+\bm{\delta}_i|$ hold.

The self consistency equations for two constant vector potentials $\bm{\delta}_i$ are reduced to 
\begin{equation}
\left\{
\begin{array}{c} 
\bm{\delta}_1=\displaystyle K\sum_{\bm{q}}
\frac{\bm{q}+\bm{\delta}_1}{|\bm{q}+\bm{\delta}_1|}
+\frac{1}{2}\frac{\bm{q}+\bm{\delta}_2}{|\bm{q}+\bm{\delta}_2|},\\
\bm{\delta}_2=\displaystyle K\sum_{\bm{q}}
\frac{1}{2}\frac{\bm{q}+\bm{\delta}_1}{|\bm{q}+\bm{\delta}_1|}
+\frac{\bm{q}+\bm{\delta}_2}{|\bm{q}+\bm{\delta}_2|}.
\end{array}
\right.
\label{SCEofDeltas}
\end{equation}
The Lorentz covariant estimation (\ref{LCE}) gives
\begin{equation}
\displaystyle K\sum_{\bm{q}}
\displaystyle\frac{\bm{q}+\bm{\delta}}{|\bm{q}+\bm{\delta}|}
=\xi^2\bm{\delta}
\left(1-\frac{4\bm{\delta}^2}{15\Lambda^2}\right).
\label{EstmOfIntegral}
\end{equation}
The form of equations (\ref{SCEofDeltas}) show that two vector potentials $\bm{\delta}_i$ are collinear and expressible in the form $\bm{\delta}_i=\delta_i\bm{e}$ with the help of the unit vector $\bm{e}$ in the common direction. 
Adding and subtracting two equations of (\ref{SCEofDeltas}) we have the equations for $\delta_i$:
\begin{equation}
\left\{
\begin{array}{l} 
(\delta_1+\delta_2)\left[\frac{2}{3\xi^2}-1
+\frac{4}{15\Lambda^2}(\delta_1^2-\delta_1\delta_2+\delta_2^2)\right]=0,\\
(\delta_1-\delta_2)\left[\frac{2}{\xi^2}-1
+\frac{4}{15\Lambda^2}(\delta_1^2+\delta_1\delta_2+\delta_2^2)\right]=0.
\end{array}
\right.
\label{SCEofSDeltas}
\end{equation}
The multitude of solutions for the algebraic equations (\ref{SCEofSDeltas}) depends on the magnitude of $\xi$. These are
\begin{equation}
\begin{array}{ll} 
{\rm G}_1:&
\delta_1=\delta_2=\pm\frac{2\pi\sqrt{15}}{g}m_A\sqrt{\xi^2-2/3},\\
{\rm G}_2:&
\delta_1=-\delta_2=\pm\frac{2\pi\sqrt{15}}{g}m_A\sqrt{\xi^2-2},\\
{\rm G}_3:&
(\delta_1,\delta_2)\;{\rm or}\;(\delta_2,\delta_1)=\left(\pm\frac{\pi\sqrt{15}}{g}m_A(2\xi+\sqrt{\xi^2-8/3}), \mp\frac{\pi\sqrt{5}}{g}m_A(2\xi-\sqrt{\xi^2-8/3})\right).
\end{array}
\label{SolofSDeltas}
\end{equation}
The number of generations increases with increasing $\xi$ as shown in Table \ref{table1}.
We have not counted the solutions obtained by reversing the signs or interchanging the names of $\delta_i$ as independent.
The maximum number of generations is 3, or equivalently, the maximum quasi-quark flavors are 6.
\begin{table}
\caption{The number of generations $G$ for quasi quarks}
\label{table1}
\begin{center}
\begin{tabular}{cccccccccc}\hline\hline
$\xi$& 0&&$\sqrt{2/3}$&&$\sqrt{2}$&&$\sqrt{8/3}$&&$+\infty$ \\ \hline
$G$&&0&&
1&&
2&&
3\\ 
\hline
\end{tabular}
\end{center}
\end{table}

The matrix elements of the Hamiltonian with respect to the one-quasi-quark states are calculated as
\begin{equation}
\langle q_aHq_b^\dagger\rangle=
\left[\begin{array}{cccc} 
|\bm{p}+\bm{\delta}_1|&0&0&0\\
0&|\bm{p}+\bm{\delta}_2|&0&0\\
0&0&|\bm{p}-\bm{\delta}_1|&0\\
0&0&0&|\bm{p}-\bm{\delta}_2|\\
\end{array}\right],
\label{Matrix_H}
\end{equation}
where $q_a$ runs over $q_{1\bm{p}}$, $q_{2\bm{p}}$, $\bar{q}_{1\bm{p}}$, and $\bar{q}_{2\bm{p}}$, which shows that the one-particle states $q_{i\bm{p}}^\dagger|{\rm VM}\rangle$  are actually the one-particle energy eigenstates of the Hamiltonian, which will contrast to the result of the next section. Each expression of energies is quasi relativistic, with the intrinsic mass $m=0$ and with the scalar (time-like) potential $\delta_0=0$.

\section{Quasi quarks of type VD}
What happens when we construct quasi-fermions from the vacuum with the Cooper pairs forming vector di-ons:
 \begin{equation}
|{\rm VD}\rangle=\prod_{\bm{p}}\left[\left(\alpha_{1\bm{p}}+\beta_{1\bm{p}}(a_{\bm{p}\LL}a_{-\bm{p}\RR})^\dagger\right)\left(\alpha_{2\bm{p}}+\beta_{2\bm{p}}(b_{\bm{p}\LL}b_{-\bm{p}\RR})^\dagger\right)\right]|0\rangle,
\label{VDV}
\end{equation}
will possibly be a question worth examined in view of Lorentz invariance of the emergent theory.
The annihilation operators of quasi fermions are, in this case, definable in the form
\begin{equation}
\begin{array}{ll}
\left\{\begin{array}{l}
q_{1\bm{p}}=\alpha_{1\bm{p}}a_{\bm{p}\LL}+\beta_{1\bm{p}}a_{-\bm{p}\RR}^\dagger,\\
q_{2\bm{p}}=\alpha_{2\bm{p}}b_{\bm{p}\LL}+\beta_{2\bm{p}}b_{-\bm{p}\RR}^\dagger,
\end{array}\right.
&
\left\{\begin{array}{l}
\bar{q}_{1\bm{p}}=\alpha_{1-\bm{p}}a_{\bm{p}\RR}-\beta_{1-\bm{p}}a_{-\bm{p}\LL}^\dagger,\\
\bar{q}_{2\bm{p}}=\alpha_{2-\bm{p}}b_{\bm{p}\RR}-\beta_{2-\bm{p}}b_{-\bm{p}\LL}^\dagger,
\end{array}\right.
\end{array}
\label{VDQ}
\end{equation}
which annihilate $|{\rm VD}\rangle$, and obey the ordinary anti-commutation rules.
The energy of the vacuum $|{\rm VD}\rangle$: 
$W=\langle{\rm VD}| H|{\rm VD}\rangle$ is estimated as 
\begin{equation}
\begin{array}{ll}
W&=\displaystyle\sum_{\bm{p}}2|\bm{p}|(|\beta_{1\bm{p}}|^2+|\beta_{2\bm{p}}|^2)\\
&+K\displaystyle\sum_{\bm{p},\bm{q}}\left[
\begin{array}{l}
-3/2\\
+\frac{1}{2}(1-3\bm{e}_{\bm{p}}\cdot\bm{e}_{\bm{q}})
(|\beta_{1\bm{p}}|^2-|\beta_{2-\bm{p}}|^2)(|\beta_{1\bm{q}}|^2-|\beta_{2-\bm{q}}|^2)\\
-\frac{1}{2}\bm{e}^+_{\bm{p}}\cdot\bm{e}^-_{\bm{q}}
(\alpha_{1\bm{p}}\beta^*_{1\bm{p}}+\alpha^*_{2-\bm{p}}\beta_{2-\bm{p}})
(\alpha^*_{1\bm{q}}\beta_{1\bm{q}}+\alpha_{2-\bm{q}}\beta^*_{2-\bm{q}})\\
\end{array}\right].
\end{array}
\label{E_VD}
\end{equation}
The variational equations (\ref{ABCD}) then give the self consistency equations for $m_{i\bm{p}}$ and $\delta_{i\bm{p}}$:
\begin{equation}
\left\{\begin{array}{ll} 
m_{1\bm{p}}=\displaystyle K\sum_{\bm{p}}
&\frac{1}{2}\bm{e}^+_{\bm{p}}\cdot\bm{e}^-_{\bm{q}}
\left(\alpha^*_{1\bm{q}}\beta_{1\bm{q}}+\alpha_{2-\bm{q}}\beta^*_{2-\bm{q}}\right),
\\
\delta_{1\bm{p}}=\displaystyle K\sum_{\bm{p}}
&\frac{1}{2}(1-3\bm{e}_{\bm{p}}\cdot\bm{e}_{\bm{q}})
\left(|\beta_{1\bm{q}}|^2-|\beta_{2-\bm{q}}|^2\right),
\end{array}\right.
\label{M1&D1(VD)}
\end{equation}
\begin{equation}
\left\{\begin{array}{ll} 
m_{2\bm{p}}=\displaystyle K\sum_{\bm{p}}
&-\frac{1}{2}\bm{e}^+_{\bm{p}}\cdot\bm{e}^+_{\bm{q}}
\left(\alpha_{1\bm{q}}\beta^*_{1\bm{q}}+\alpha^*_{2-\bm{q}}\beta_{2-\bm{q}}\right),
\\
\delta_{2\bm{p}}=\displaystyle K\sum_{\bm{p}}
&-\frac{1}{2}(1+3\bm{e}_{\bm{p}}\cdot\bm{e}_{\bm{q}})
\left(|\beta_{1\bm{q}}|^2-|\beta_{2-\bm{q}}|^2\right).
\end{array}\right.
\label{M2&D2(VD)}
\end{equation}
In contrast to the previous case, we find $m_{2\bm{p}}=m_{1-\bm{p}}^*$, $\delta_{2\bm{p}}=-\delta_{1-\bm{p}}$, and further that $m_{i\bm{p}}$ and $\delta_{i\bm{p}}$ are expressible as 
\begin{equation}
\begin{array}{cc} 
\left\{\begin{array}{l} 
m_{1\bm{p}}=\bm{e}^+_{\bm{p}}\cdot\bm{m},\\
\delta_{1\bm{p}}=\delta_0+\bm{e}_{\bm{p}}\cdot\bm{\delta},
\end{array}\right.
&
\left\{\begin{array}{l} 
m_{2\bm{p}}=-\bm{e}^+_{\bm{p}}\cdot\bm{m},\\
\delta_{2\bm{p}}=-\delta_0+\bm{e}_{\bm{p}}\cdot\bm{\delta},
\end{array}\right.
\end{array}
\label{M&D(VD)}
\end{equation}
where two vectors $\bm{m}$, $\bm{\delta}$, and a scalar 
$\delta_0$ are real and constant. Then (\ref{M1&D1(VD)}) and (\ref{M2&D2(VD)}) give the self consistency equations  
\begin{subequations}
\begin{eqnarray}
\bm{m}&=\displaystyle\frac{K}{2}&\sum_{\bm{q}}
\bm{e}^-_{\bm{q}}\left(\alpha^*_{1\bm{q}}\beta_{1\bm{q}}+\alpha_{2-\bm{q}}\beta^*_{2-\bm{q}}\right),\label{SCEa(VD)}\\
\bm{\delta}&=-\displaystyle\frac{3}{2}K\displaystyle&\sum_{\bm{q}}
\bm{e}_{\bm{q}}\left(|\beta_{1\bm{q}}|^2-|\beta_{2-\bm{q}}|^2\right).
\label{SCEb(VD)}\\
\delta_0&=\displaystyle\frac{K}{2}\displaystyle&\sum_{\bm{q}}
\left(|\beta_{1\bm{q}}|^2-|\beta_{2-\bm{q}}|^2\right).\label{SCEc(VD)}
\end{eqnarray}
\end{subequations}
We see that quasi fermions of type VD are generally not quasi relativistic. In order for that, we should impose the conditions $\bm{m}=\pm\bm{\delta}$ with $\delta_0=0$ in contrast to the case of type VM. At first sight, the case $\bm{m}=\bm{\delta}=0$ with $\delta_0\neq0$ would give also quasi relativistic fermions, but this is not true, since then the dispersion relation would be $\omega=||\bm{p}|\pm\delta_0|$, which does not belong to the class (\ref{QRDR}). In addition, it is confirmed that this case is inconsistent with (\ref{SCEc(VD)}), since then the right hand side has the opposite sign to the left.

We proceed to examine the self consistency equations (\ref{SCEa(VD)}) and (\ref{SCEb(VD)}) with assuming $\delta_0=0$, but without $\bm{m}=\pm\bm{\delta}$.
Then we have
\begin{subequations}
\begin{eqnarray}
\bm{m}&=\displaystyle\frac{K}{2}&\sum_{\bm{q}}
\frac{\bm{m}-\bm{e}_{\bm{q}}(\bm{e}_{\bm{q}}\cdot\bm{m})}{\omega_q},\label{SCEm(VD)}\\
\bm{\delta}&=\displaystyle\frac{3}{2}K&\sum_{\bm{q}}
\frac{\bm{q}+\bm{e}_{\bm{q}}(\bm{e}_{\bm{q}}\cdot\bm{\delta})}{\omega_q},
\label{SCEd(VD)}
\end{eqnarray}
\end{subequations}
where $\omega_{\bm{q}}=\sqrt{(|\bm{q}|+\bm{e}_{\bm{q}}\cdot\bm{\delta})^2+|\bm{e}^+_{\bm{q}}\cdot\bm{m}|^2}$.
The mass vector $\bm{m}$ and the vector potential $\bm{\delta}$ are now not necessarily collinear, in contrast to the case VM again. We further assume $\bm{m}\times\bm{\delta}=0$, and introduce the scalar parameters ($m, \delta)$ by $\bm{m}=m\bm{e}$ and $\bm{\delta}=\delta\bm{e}$ with the help of an appropriate unit vector $\bm{e}$.
Since the equations (\ref{SCEm(VD)}) and (\ref{SCEd(VD)}) are invariant under the reflections $\bm{m}\leftrightarrow-\bm{m}$ and/or $\bm{\delta}\leftrightarrow-\bm{\delta}$, we henceforth assume $m>0$ and $\delta>0$ for simplicity.

The momentum summations are analytically estimable, though  their Lorentz covariant forms can not be found. Introducing dimensionless parameters $x=m/\Lambda$ and $y=\delta/\Lambda$, we have for (\ref{SCEm(VD)}) and (\ref{SCEd(VD)}),
\begin{equation}
\left\{
\begin{array}{cl} 
\bm{m}=&\bm{m}\xi^2F(x,y),\\
\bm{\delta}=&\bm{\delta}\xi^2G(x,y),
\end{array}\right.
\label{F&G}
\end{equation}
where
\begin{equation}
F(x,y)=\left\{
\begin{array}{cr} 
\quad\displaystyle\frac{2}{15}(y^2-2w^2)\ln\frac{y^2-w^2}{1-w^2}
+\frac{4}{15}-\frac{1}{5w^2}+y^2(-\frac{4}{15}-\frac{2}{15w^2}+\frac{3}{5w^4})&\\
+\left\{\displaystyle\frac{1}{3}+\frac{1}{5w^2}+y^2(\frac{1}{3w^2}-\frac{3}{5w^4})\right\}\displaystyle\frac{1}{2w}\ln\frac{1+w}{1-w},&( y\leq1),\\
&\\
 y(-\displaystyle\frac{1}{3w^2}+\frac{3}{5w^4})
+\left\{\frac{1}{3}+\frac{1}{5w^2}+y^2(\frac{1}{3w^2}-\frac{3}{5w^4})\right\}\frac{1}{2w}\ln\frac{1+w/y}{1-w/y},&( y>1).
\end{array}\right.
\label{F(x,y)}
\end{equation}
and
\begin{equation}
G(x,y)=\left\{
\begin{array}{cr} 
\quad-\displaystyle\frac{2}{5}(y^2-w^2)
\ln\frac{y^2-w^2}{1-w^2}
-\frac{2}{5}+\frac{9}{5w^2}+y^2(\frac{2}{5}+\frac{2}{5w^2}-\frac{9}{5w^4})&\\
+\left\{1-\displaystyle\frac{9}{5w^2}+y^2(-\frac{1}{w^2}+\frac{9}{5w^4})\right\}\displaystyle\frac{1}{2w}\ln\frac{1+w}{1-w},&( y\leq1),\\
&\\
\displaystyle\frac{6}{5w^2y}+
y(\frac{1}{w^2}-\frac{9}{5w^4})
+\left\{1-\frac{9}{5w^2}
+y^2(-\frac{1}{w^2}+\frac{9}{5w^4})\right\}\frac{1}{2w}\ln\frac{1+w/y}{1-w/y},&( y>1),
\end{array}\right.
\label{G(x,y)}
\end{equation}
with $w=\sqrt{y^2-x^2}$. The values of functions $F$ and $G$ for $x>y$ are obtainable by the analytic continuation without any ambiguity.

The non-trivial solutions for (\ref{F&G}) are classified into two singular cases; S$_1$: $\bm{m}\neq0$ with $\bm{\delta}=0$, S$_2$: $\bm{m}=0$ with $\bm{\delta}\neq0$, and one regular case RG: $\bm{m}\neq0$ with $\bm{\delta}\neq0$.
The self consistency equation for the case S$_1$ reads 
\begin{equation}
\xi^{-2}=F(x,0)=\frac{4}{15}\left\{1-x^2\ln(1+\frac{1}{x^2})\right\}+\frac{1}{5x^2}+(\frac{1}{3x}-\frac{1}{5x^3})
\arcsin\frac{x}{\sqrt{x^2+1}}.
\label{F(x,0)}
\end{equation}
The function $F(x,0)$ is monotonically decreasing and $F(0,0)=2/3$. Then (\ref{F(x,0)}) has one solution for $\xi>\sqrt{3/2}$.

The self consistency equation for the case S$_2$ reads 
\begin{equation}
\xi^{-2}=G(0,y)=
\left\{
\begin{array}{lr} 
\displaystyle\frac{2}{5}y^2&(y\leq1),\\
\displaystyle\frac{1}{y}-\frac{3}{5y^3}&(y>1).
\end{array}\right.
\label{G(0,y)}
\end{equation}
The function $G(0,y)$ is monotonically increasing for $y<1$,   decreasing for  $y>1$, and attains its maximum $2/5$ at $y=1$. Then (\ref{G(0,y)}) has two solutions for $\xi>\sqrt{5/2}$.

\begin{figure}
\includegraphics[width=3in]{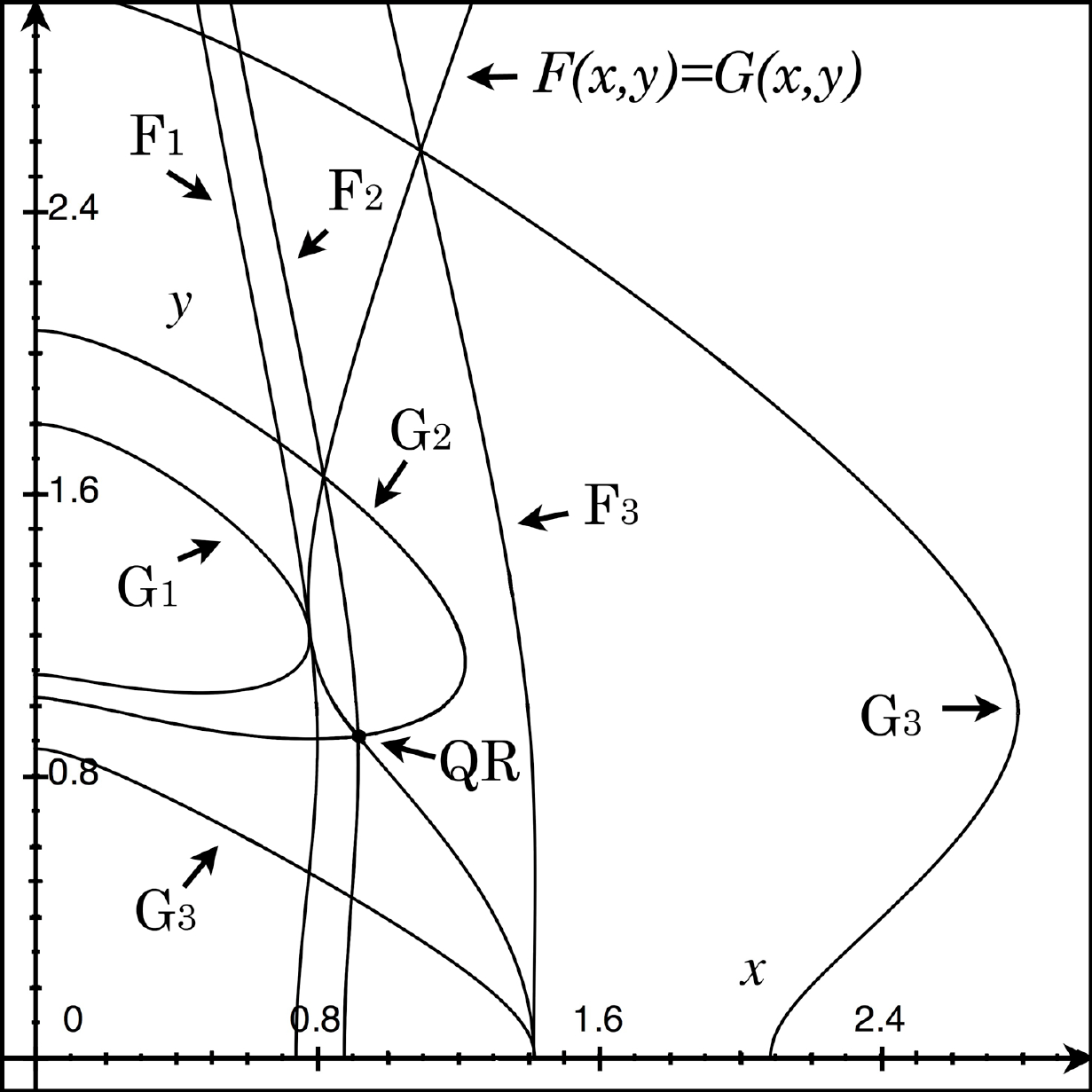}
\caption{The graphical representation of the regular solutions for quasi-quarks of type VD.  F$_{1,2,3}$ and G$_{1,2,3}$ denote the curves $F(x,y)=\xi^{-2}$ and $G(x,y)=\xi^{-2}$ with varying $\xi$, respectively. F$_1$ and G$_1$ are at $\xi=1.486$, F$_2$ and G$_2$ are at $\xi=1.55$, and F$_3$ and G$_3$ are at $\xi=1.802$. The curve $F(x,y)=G(x,y)$ shows the mass-potential correlation. QR is the quasi relativistic point.}
\label{XY}
\end{figure}

The solutions for the remaining case RG will be difficult to find  without any computer software, which shows no solution for $\xi<\xi_1\simeq1.486$, two solutions for $\xi_1<\xi<\xi_2\simeq1.802$ and one solution for $\xi_2<\xi$ as seen from  Fig.\ref{XY}.
Table \ref{table2} summarizes the total number of generations for the quasi quarks of type VD. 
\begin{table}
\caption{The number of generations $G$ for quasi quarks of type VD}
\label{table2}
\begin{center}
\begin{tabular}{cccccccccccc}\hline\hline
$\xi$& 0&&$\sqrt{3/2}$&&$\xi_1$&&$\sqrt{5/2}$&&$\xi_2$&&$+\infty$ \\ \hline
$G$&&0&&
1&&
3&&
5&&4\\ 
\hline
\end{tabular}
\end{center}
\end{table}

The quasi quarks of type VD are generally not quasi relativistic. The quasi relativistic quarks emerge only when $\xi=\xi_{\rm QR}\simeq1.55$ with $x=y\simeq0.9115$. Table \ref{table2} shows that $\xi_{\rm QR}$ lies in the region of 5 generations. Fig.\ref{XY} indicates, however, that even in this case, only one generation can be quasi relativistic.

The matrix elements of the Hamiltonian between one-quasi-quark states of type VD have the following form
\begin{equation}
\langle q_aHq_b^\dagger\rangle_{\rm VD}=
\left[\begin{array}{cccc} 
\omega_{\bm{p}}&i\gamma_{\bm{p}}&0&0\\
-i\gamma_{\bm{p}}&\omega_{\bm{p}}&0&0\\
0&0&\bar{\omega}_{\bm{p}}&i\bar{\gamma}_{\bm{p}}\\
0&0&-i\bar{\gamma}_{\bm{p}}&\bar{\omega}_{\bm{p}}\\
\end{array}\right],
\label{Matrix_H(VD)}
\end{equation}
where $\bar{\omega}_{\bm{p}}=\omega_{-\bm{p}}$, $\bar{\gamma}_{\bm{p}}=\gamma_{-\bm{p}}$ and
\begin{equation}
\gamma_{\bm{p}}=\frac{\bm{m}\cdot(\bm{p}+\bm{\delta})}{\omega_{\bm{p}}}.
\label{gamma(VD)}
\end{equation}
One remarkable feature for quasi quarks of type VD is that they are not in energy eigenstates.
The off diagonal elements $\gamma_{\bm{p}}$ and $\bar{\gamma}_{\bm{p}}$ in (\ref{Matrix_H(VD)}) seem to generate phenomena similar to the quark mixings \cite{Cabibbo,GIM,CKM} and the neutral meson oscillations \cite{CCFT}.
The energy eigenstates $|Q^+_{\bm{p}}\rangle=(|q_{1\bm{p}}\rangle)-i|q_{2\bm{p}}\rangle)/\sqrt{2}$ and $|Q^-_{\bm{p}}\rangle=(-i|q_{1\bm{p}}\rangle+|q_{2\bm{p}}\rangle)/\sqrt{2}$ have energy eigenvalues $\omega_{\bm{p}}\pm\gamma_{\bm{p}}$.  The mixing angle between $q_1$ and $q_2$ quasi-quarks is therefore $\pi/4$, and they oscillate with the probability
\begin{equation}
|\langle q_{2\bm{p}}|q_{1\bm{p}}(t)\rangle|^2=\sin^2\gamma_{\bm{p}} t.
\label{Quark Oscillation}
\end{equation}
In contrast to the real observations, however, mixings and oscillations occur in the same doublet, and $\gamma_{\bm{p}}$ is generally of the same order of the quark mass parameter, which could make the frequency of oscillation very high. Though the properties of quasi quarks of type VD are not suitable for  describing real phenomena, the mixing phenomenon by spontaneous space-time symmetry violation may shed some light on the quark or neutrino mixings.

Incidentally, the analysis in this section does not necessarily prove the emergence of non-relativistic quasi fermions in our model. In order for that, the vacuum $\vert{\rm  VD}\rangle$ should be, at least, local minimum in the whole configuration space of the vacuum, and further energetically more favorable than $\vert{\rm  VM}\rangle$.
Even for that case, it would be unclear whether the non-relativistic quasi fermions in confinement phase immediately contradict to Lorentz symmetry, until the relativistic dispersion relation of a single quark can be experimentally confirmed.

\section{Summary}
We have shown in this paper that a unified picture of fermions, in which the vacuum expectation values of SU(2) potentials are assumed to give a mass term to a chiral doublet, allows two and only two types of quasi-relativistic quasi-fermions. 
This distinction seems to reflect that between leptons and quarks in reality.

New quasi-fermion doublets are massless and anisotropic particles which have constant vector potentials in dispersion relations.
They seem to qualify for the name ``quasi-quarks", since the first paper suggests that the anisotropy may be the origin of color degrees of freedom.

Quasi-quarks in a doublet derived by a perturbative method have vector potentials with the same magnitude and opposite signs, while those by a variational method can have potentials with different magnitudes. 

The variational method also reproduces a generation structure similar to that of real quarks.
While the perturbative method derives only one generation, the variational method can derive maximally three generations.

The validity of the analysis based on a four fermion approximation of the original theory (\ref{Ham}) seems to restrict the energy scale to be lower than $m_A$. If this upper limit applies to the cutoff scale $\Lambda$, then $\xi$ should be nearly less than $1/20$. However, we have seen that quasi quarks require larger values, $\xi\simeq1$, which may give rise to a question for the validity of the results obtained here.  We have seen in the first paper that the quasi leptons require $\xi>1$. Therefore, the value $\xi\simeq1$ required for quasi quarks seems to indicate the  characteristics of the original theory, not influenced by the four fermion approximation.

The aim of this paper is simply to demonstrate the emergence of  a special type of quasi-fermions, which are expected but not proved in the first paper for the quasi-fermion representation of quarks. To discuss on the equivalence between quasi-quarks and the quarks in QCD \cite{FGL}, and to strengthen such a view by providing with more arguments or evidences are not within the scope of this paper. For that purpose, many problems remain yet to be solved.

\appendix
\section{Lorentz covariant estimation of a divergent integral}
The estimation of divergent integrals have some subtleties and the results depend on the ways of estimation.
For example, the Lorentz covariant estimation of a quadratically  divergent integral 
\begin{equation}
\displaystyle\int\frac{d^4p}{(2\pi)^4}i\frac{p^\mu p^\nu}{(p^2+i\epsilon)^2},
\end{equation}
gives $g^{\mu\nu}k_1/4$, while estimating it by doing the time-like integration first gives $(g^{\mu\nu}+2g^{\mu0}g^{\nu0})k_1/6$. Accordingly, the Lorentz covariant estimation of $S(0)$ in (\ref{S(0)}) requires some cautions. We first simplify $S(0)$ by a unitary transformation $U_\epsilon\bm{\rho}\cdot\bm{\epsilon}U^{-1}_\epsilon=\rho_3$: 
\begin{equation}
\begin{array}{cc} 
U_{\epsilon} S(0)U^{-1}_\epsilon=\displaystyle\frac{1+\rho_3}{2}s_-(0)+\frac{1-\rho_3}{2}s_+(0),&
s_\pm(0)=\displaystyle\int\frac{d^4p}{(2\pi)^4}
\frac{i}{\bar{\sigma}\cdot(p\pm\delta)}=\sigma\cdot I_\pm
\end{array}
\label{U-transf}
\end{equation}
where
\begin{equation}
I^\mu_\pm=\displaystyle\int\frac{d^4p}{(2\pi)^4}i
\frac{p^\mu\pm\delta^\mu}{(p\pm\delta)^2}.
\label{Imu}
\end{equation}
The Taylor expansion of (\ref{Imu}) with respect to $\delta^\mu$ gives
\begin{equation}
I^\mu_\pm=\pm\displaystyle\int\frac{d^4p}{(2\pi)^4}i
\left[\frac{g^{\mu\nu}}{p^2+i\epsilon}-2\frac{p^\mu p^\nu}{(p^2+i\epsilon)^2}\right]\delta_\nu+{\cal O}(\delta^3)
=\pm k_1\delta^\mu/2+{\cal O}(\delta^3),
\label{TexImu}
\end{equation}
in the Lorentz covariant estimation, while doing the time-like integration first gives
\begin{equation}
\bm{I}_\pm=\frac{1}{2}\displaystyle\int\frac{d^3p}{(2\pi)^3}
\frac{\bm{p}\pm\bm{\delta}}{|\bm{p}\pm\bm{\delta}|}=
\pm\bm{\delta}\left[\frac{2k_1}{3}-\frac{\bm{\delta}^2}{60\pi^2}\right].
\end{equation}
Therefore, we obtain  
\begin{equation}
\bm{I}_\pm^{({\rm LCE})}=
\pm\bm{\delta}\left[\frac{k_1}{2}-\frac{\bm{\delta}^2}{60\pi^2}\right],
\label{LCE}
\end{equation}
in the Lorentz covariant estimation, since the ${\cal O}(\delta^3)$ part is finite and independent of the ways of estimation. Consequently, we have 
\begin{equation}
\begin{array}{cc} 
U_{\epsilon} S(0)U^{-1}_\epsilon=\rho_3s_-(0),&
s_\pm(0)=\pm\sigma\cdot\delta
\left[\displaystyle\frac{k_1}{2}+\frac{\delta\cdot\delta}{60\pi^2}\right].
\end{array}
\end{equation}
Finally, the inverse transformation of 
(\ref{U-transf}) gives the result (\ref{S-estimation}).


\begin{thebibliography}{}
\bibitem{KN1}
K. Nishimura, Prog. Theor. Exp. Phys. 023B06 (2013); doi: 10.1093/ptep/pts091.
\bibitem{KN2}
K. Nishimura, arXiv:1211.1038.
\bibitem{Sakhalov}
A. D. Sakhalov, JETP Lett. {\bf 5}, 24 (1967).
\bibitem{Trodden}
M. Trodden, Rev. Mod. Phys. {\bf 71}, 1463 (1999).
\bibitem{KT}
E. W. Kolb and M. S. Turner, ``The Early Universe", Chap.3, Sec. 4 and 5, Addison-Wesley (1990).
\bibitem{CK}
D. Colladay and V. A. Kosteleck\'{y}, Phys. Rev. D55  6760 (1997) 
\bibitem{BCS}
J. Bardeen, N. Cooper and R. Schriefer, Phys. Rev. {\bf108},1175 (1957).
\bibitem{Cabibbo}
N. Cabibbo, Phys. Rev. Lett. {\bf 10}, 531 (1963).
\bibitem{GIM}
S. L. Glashow, J. Iliopoulos and L. Maiani, Phys. Rev. {\bf D2}, 1285 (1970)
\bibitem{CKM}
M. Kobayashi, T. Maskawa, Prog.Theor. Phys. {\bf 49}, 652 (1973).
\bibitem{CCFT}
J.H. Christenson, J.W. Cronin, V.L. Fitch, R. Turlay, Phys. Rev. Lett. {\bf 13}, 138 (1964).
\bibitem{FGL}
H. Fritzsch, M. Gell-Mann, and H. Leutwyler, Phys. Lett. {\bf 47B}, 365 (1973).
\end{thebibliography}
\end{document}